\documentclass[aps,preprint,amsmath,tightenlines]{revtex4}

\usepackage{bm}

\begin{document}

\title{Bounds on quark mass matrices elements\\ due to measured
properties of the mixing matrix\\ and present values of the quark masses}

\author{S.~Chaturvedi}

\email[]{scsp@uohyd.ernet.in}

\affiliation{
School of Physics.\\
University of Hyderabad\\
Hyderabad 500 046, India.
}

\author{V.~Gupta}

\email[]{virendra@mda.cinvestav.mx}

\affiliation{
Departamento de F\'{\i}sica Aplicada.\\
Centro de Investigaci\'on y de Estudios Avanzados del IPN.\\
Unidad Merida.\\
A.P. 73, Cordemex.\\
M\'erida, Yucat\'an, 97310. MEXICO.
}

\author{G.~S\'anchez-Col\'on}

\email[]{gsanchez@mda.cinvestav.mx}

\altaffiliation[Permanent address: ]{Departamento de F\'{\i}sica
Aplicada.\\ CINVESTAV del IPN.
Unidad Merida. A.P. 73, Cordemex. M\'erida, Yucat\'an, 97310.
MEXICO.}

\affiliation{
Facultad de Matem\'aticas.\\
Universidad Autonoma de Yucat\'an.\\
Anillo Periferico Norte. Tablaje Cat. 13615, Chuburna Hidalgo Inn.\\
M\'erida, Yucat\'an, 97203. MEXICO.
}

\date{\today}

\begin{abstract}

We obtain constraints on possible structures of mass matrices in the
quark sector by using as experimental restrictions the
determined values of the quark masses at the $M_Z$~energy scale,
the magnitudes of the quark mixing matrix elements $V_{\rm ud}$,
$V_{\rm us}$, $V_{\rm cd}$, and $V_{\rm cs}$, and the Jarlskog
invariant $J(V)$. Different cases of specific mass matrices are
examined in detail. The quality of the fits for the Fritzsch and
Stech type mass matrices is about the same with $\chi^2/{\rm
dof}=4.23/3=1.41$ and $\chi^2/{\rm dof}=9.10/4=2.28$,
respectively. The fit for a simple generalization (one extra
parameter) of the Fritzsch type matrices, in the physical basis,
is much better with $\chi^2/{\rm dof}=1.89/4=0.47$. For
comparison we also include the results using the quark masses at
the 2~GeV~energy scale. The fits obtained at this energy scale
are similar to that at $M_Z$ energy scale, implying that our
results are unaffected by the evolution of the quark masses from
2 to 91~GeV.

\end{abstract}


\maketitle

\section{\label{introduction}Introduction}

Quark flavor mixing in the Standard Model arises from unitary matrices
which diagonalize the corresponding hermitian mass matrices. The CKM
quark-flavor mixing matrix $V$ in the physical basis~\cite{2a,2b,2c}, is
given by $V= V_{\rm u}V^\dagger_{\rm d}$, where the unitary matrices
$V_{\rm u}$ and $V_{\rm d}$ diagonalize the up-quark and down-quark
Dirac mass matrices $M_{\rm u}$ and $M_{\rm d}$, respectively. Given
this circumstance, complete knowledge of the mass matrices fully
determines the corresponding mixing matrices. In practice, however, the
mass matrices are guessed at, while experiment can only determine the
moduli of the CKM matrix elements.

Recently, evolved quark masses have been given at various energy
scales~\cite{xing07}. The quark masses at the
$M_W\,(=80.403\,{\rm GeV})$, $M_Z\,(=91.1876\,{\rm GeV})$, and
$m_{\rm t}\,(=172.5\,{\rm GeV})$ scales are quite similar. For
our analysis we use as input the quark masses at the
$M_Z$~energy scale, which lies between the $M_W$ and the $m_t$
scales. So, in this paper we shall use as experimental
restrictions the reported values of the CKM matrix
elements $|V_{\alpha\, j}|$~\cite{pdg06}, the Jarlskog invariant
$J(V)$~\cite{pdg06}, and the quark masses at the $M_Z$ energy
scale~\cite{xing07}, to obtain constraints on the elements of
the quark mass matrices in five specific cases for three
generations. As a check of the stability of our type of
analysis, under the evolution of quark masses, we have repeated
it with quark masses at 2~GeV scale~\cite{xing07}.

Section~\ref{notation} gives the notation and the basic formulas or
expressions needed and the general procedure adopted for the analysis.
In Sec.~\ref{fritzsch}, we consider the Fritzsch-type of mass
matrices~\cite{fritzsch70,fritzsch78,fritzsch86} in the physical basis.
As pointed out there $M_{\rm u}$ can be chosen to be real with three
parameters while $M_{\rm d}$ has five, two of which are phase angles. In
Sec.~\ref{stech}, Stech-type matrices~\cite{stech83} are considered in
the basis in which $M_{\rm u}$ is diagonal, while $M_{\rm d}=p\,M_{\rm
u}+i\,S$. Here $p$ is a real number and $S$ is a non-diagonal matrix,
satisfying $S^{\rm T}=-S$, with three real parameters. In
Secs.~\ref{diagonal} and \ref{nondiagonal} we consider mass matrices
which are a simple generalization of the Fritzsch-type matrix in that
$M_{\rm u}$ and $M_{\rm d}$ have an additional parameter. These were
considered recently, in the $M_{\rm u}$ ($M_{\rm d}$) diagonal
basis~\cite{chaturvedi07}. In Sec.~\ref{diagonal} we consider these
cases again and fit them to the data. In Sec.~\ref{nondiagonal} we
consider the case of these type of mass matrices in the physical basis.
Results obtained in the above cases are compared and discussed in the
final section~\ref{conclusions}. Here also the results of the two
energy scales ($M_Z$ and 2~GeV) are compared and discussed.

\section{\label{notation}Notation and basic formulas}

The $3\times 3$ hermitian quark mass matrix $M_{\rm q}$ is diagonalized
by $V_{\rm q}$ so that $M_{\rm q}=V_{\rm q}^{\dagger}\hat{M_{\rm
q}}V_{\rm q}\,$, q=u,d. The eigenvalues are denoted by ($\lambda_{\rm
u}$,$\lambda_{\rm c}$,$\lambda_{\rm t}$) and ($\lambda_{\rm
d}$,$\lambda_{\rm s}$,$\lambda_{\rm b}$) for the up and down quark mass
matrices, respectively. Note that the eigenvalues are real but not
necessarily positive. Each mass matrix can be expressed in terms of its
projectors. Thus,

\begin{equation}
M_{\rm u}=\sum_{\alpha={\rm u, c, t}}\lambda_{\alpha}N_{\alpha}
\qquad
{\rm and}
\qquad
M_{\rm d}=\sum_{j={\rm d, s, b}}\lambda_{j}N_{j}.
\label{projectors}
\end{equation}

\noindent
Since $V= V_{\rm u}V^\dagger_{\rm d}$, it follows that~\cite{18}

\begin{equation}
|V_{\alpha\, j}|^2={\rm Tr}[N_\alpha N_j],
\label{5}
\end{equation}

\noindent
where

\begin{equation}
N_\alpha= \frac{(\lambda_\beta-M_{\rm u})(\lambda_\gamma
-M_{\rm u})}{(\lambda_\beta-\lambda_\alpha) (\lambda_\gamma
-\lambda_\alpha)}
\label{4}
\end{equation}

\noindent
and

\begin{equation}
N_j=
\frac{(\lambda_k-M_{\rm d})(\lambda_l -M_{\rm
d})}{(\lambda_k-\lambda_j) (\lambda_l -\lambda_j)},
\label{4p}
\end{equation}

\noindent
with ($\alpha$,$\beta$,$\gamma$) and ($j$,$k$,$l$) any permutation of
(u,c,t) and (d,s,b), respectively.

The Jarlskog invariant $J(V)$, which is a measure of CP-violation can be
directly expressed in terms of $M_{\rm u}$ and $M_{\rm d}$ and their
eigenvalues~\cite{4}, thus

\begin{equation}
{\rm Det}([M_{\rm u},M_{\rm d}])=2i\,D(\lambda_\alpha)D(\lambda_j) J(V),
\label{12}
\end{equation}

\noindent
where

\begin{equation}
D(\lambda_\alpha)=
(\lambda_{\rm c}-\lambda_{\rm u})(\lambda_{\rm t}-\lambda_{\rm u})
(\lambda_{\rm t}-\lambda_{\rm c})
\label{8}
\end{equation}

\noindent
and

\begin{equation}
D(\lambda_j)=
(\lambda_{\rm s}-\lambda_{\rm d})(\lambda_{\rm b}-\lambda_{\rm d})
(\lambda_{\rm b}-\lambda_{\rm s}).
\label{8p}
\end{equation}

The bases when $M_{\rm u}$ or $M_{\rm d}$ is diagonal are of special
interest for the mass matrices considered in Secs.~\ref{stech} and
\ref{diagonal}. For the case, $M_{\rm u}$ diagonal and $M_{\rm d}=M$,
Eq.~(\ref{12}) reduces to~\cite{chaturvedi07}

\begin{equation}
J(V)=
\frac{{\rm Im}(M_{12}\,M_{23}\,M^{*}_{13})}{D(\lambda_j)}.
\label{jdiag}
\end{equation}

\noindent
There is a similar formula for the case $M_{\rm d}$ diagonal.
This result shows that to obtain CP-violation, the mass matrix for
up-quark (down-quark) must have ${\rm
Im}(M_{12}\,M_{23}\,M^{*}_{13})$ non-zero in a basis in which the
down-quark (up-quark) mass matrix is diagonal. Consequently, the
Fritzsch type of mass matrices can only be used in the physical basis.

In general, our procedure in each case is to first determine the
elements of the quark mass matrices in terms of the
eigenvalues and then to form a $\chi^2$-function which contains
eleven summands. The first five compare the theoretical
expressions as functions of the elements of the quark mass
matrices of the four best measured moduli, namely, $|V_{\rm
ud}|$, $|V_{\rm us}|$, $|V_{\rm cd}|$, and $|V_{\rm cs}|$, and of
the Jarlskog invariant $J(V)$, with their experimental
values~\cite{pdg06}. The last six summands constrain the quark
mass matrices eigenvalues to the experimentally deduced quark
masses values at the $M_Z$ energy scale~\cite{xing07}.

\section{\label{fritzsch}Fritzsch type mass matrices}

We consider first the well-known case of the Fritzsch mass
matrices~\cite{fritzsch70,fritzsch78,fritzsch86}, given by the hermitian
matrices

\begin{equation}
M_{\rm u}=
\begin{pmatrix}
0 & A & 0 \\
A^* & 0 & B \\
0 & B^* & C
\end{pmatrix}
,
\quad\quad
M_{\rm d}=
\begin{pmatrix}
0 & A' & 0 \\
A'^* & 0 & B' \\
0 & B'^* & C'
\end{pmatrix}
.
\label{13}
\end{equation}

\noindent
Without lack of generality we may take $C$ and $C'$ to be positive. Since
we may rotate $M_{\rm u}$ and $M_{\rm d}$ with the same unitary matrix $X$
without changing the physics~\cite{Jarlskog87,Jarlskog87b}, we can make
$M_{\rm u}$ real with positive elements by choosing

\begin{equation}
X=
\begin{pmatrix}
e^{-i\phi_A} & 0 & 0 \\
0 & 1 & 0 \\
0 & 0 & e^{i\phi_B}
\end{pmatrix}
,
\label{13p}
\end{equation}

\noindent
with $\phi_A$ and $\phi_B$ the phases of $A$ and $B$, respectively. Then,
$M_{\rm u}$ and $M_{\rm d}$ have eight parameters, $A$, $B$, $C$, $|A'|$,
$|B'|$, $C'$, and the phases $\phi_{A'}$ and $\phi_{B'}$.

From the characteristic equation of $M_{\rm u}$, we have

\[
C=\lambda_{\rm u}+\lambda_{\rm c}+\lambda_{\rm t},
\]

\begin{equation}
-A^2-B^2=\lambda_{\rm u}\lambda_{\rm c} +\lambda_{\rm u}\lambda_{\rm
t}+\lambda_{\rm c}\lambda_{\rm t},
\label{15}
\end{equation}

\[
-A^2C=\lambda_{\rm u}\lambda_{\rm c}\lambda_{\rm t}.
\]

\noindent
Solving for $A$ and $B$ yields

\[
A=\left[
-\frac{\lambda_{\rm u}\lambda_{\rm c}\lambda_{\rm t}}
{\lambda_{\rm u}+\lambda_{\rm c}+\lambda_{\rm t}
}
\right]^{1/2}
,
\]

\begin{equation}
B=\left[-
\frac{
(\lambda_{\rm t}+\lambda_{\rm c})
(\lambda_{\rm t}+\lambda_{\rm u})
(\lambda_{\rm c}+\lambda_{\rm u})}
{\lambda_{\rm u}+\lambda_{\rm c}+\lambda_{\rm t}
}
\right]^{1/2}
.
\label{15b}
\end{equation}

\noindent Similarly for $M_{\rm d}$, the parameters $|A'|$, $|B'|$, and
$C'$ are obtained from~(\ref{15}) and (\ref{15b}) by replacing
($A$,$B$,$C$) by ($|A'|$,$|B'|$,$C'$) and ($\lambda_{\rm
u}$,$\lambda_{\rm c}$,$\lambda_{\rm t}$) by ($\lambda_{\rm
d}$,$\lambda_{\rm s}$,$\lambda_{\rm b}$).

According to~(\ref{5}), the magnitudes of the unitary quark
mixing matrix elements are given in the Fritzsch case by

\begin{eqnarray}
|V_{\alpha\,j}|^2&=&
\big{[}
(\lambda_{\alpha} - \lambda_{\beta})
(\lambda_{\alpha} - \lambda_{\gamma})
(\lambda_j - \lambda_k)
(\lambda_j - \lambda_l)
\big{]}^{-1}\times
\nonumber
\\
&\ &
\bigg\{
\big{(}\lambda_{\beta}\lambda_{\gamma} + A^2 + B^2\big{)}
\big{(}\lambda_k\lambda_l + {|A'|}^2 + {|B'|}^2\big{)}
+
\big{(}\lambda_{\beta}\lambda_{\gamma} + A^2\big{)}
\big{(}\lambda_k\lambda_l + {|A'|}^2\big{)}
\nonumber
\\
&\ &+\,
\Big{[}
\left(\lambda_{\alpha} + \lambda_{\beta}\right)
\left(\lambda_{\alpha} + \lambda_{\gamma}\right)
+ B^2
\Big{]}
\left[
\left(\lambda_j + \lambda_k\right)
\left(\lambda_j + \lambda_l\right)
+ {|B'|}^2
\right]
\nonumber
\\
&\ &+\,
2
\left(\lambda_{\beta} + \lambda_{\gamma}\right)
\left(\lambda_k + \lambda_l)\,
A{|A'|}\cos(\phi_{A'}\right)
\nonumber
\\
&\ &+\,
2\lambda_{\alpha}\lambda_j\,B{|B'|}\cos(\phi_{B'}) +
2B\,A{|B'|}{|A'|}\cos(\phi_{A'} + \phi_{B'})
\bigg\}\,,
\label{vaj}
\end{eqnarray}

\noindent
where ($\alpha$,$\beta$,$\gamma$) is any permutation of (u,c,t) and
($j$,$k$,$l$) any permutation of (d,s,b). By unitarity only four of
the nine $|V_{\alpha\,j}|^2$ are independent. As mentioned in
Sec.~\ref{introduction}, we shall use the four best experimentally
measured magnitudes $|V_{\rm ud}|$, $|V_{\rm us}|$, $|V_{\rm cd}|$, and
$|V_{\rm cs}|$\,.

Finally, the Jarslkog invariant $J(V)$ given by Eq.~(\ref{12}) translates
for the Fritzsch case into,

\begin{eqnarray}
J(V)&=&
-\big{[}
(\lambda_{\rm t} - \lambda_{\rm c}) (\lambda_{\rm t} -
\lambda_{\rm u}) (\lambda_{\rm c} - \lambda_{\rm u})(\lambda_{\rm b} -
\lambda_{\rm s}) (\lambda_{\rm b} - \lambda_{\rm d})(\lambda_{\rm s} -
\lambda_{\rm d})
\big{]}^{-1} \times
\nonumber
\\ &\ &
\bigg\{
\Big[
B\,{|B'|}\sin(\phi_{B'}) - A\,{|A'|}\sin(\phi_{A'})
\Big]
\nonumber
\\ &\ & \quad \times
\Big[
A^2{|B'|}^2+B^2 {|A'|}^2-2\,A\,B\,{|A'|}\,{|B'|}\cos(\phi_{A'}+\phi_{B'})
\Big]
\nonumber
\\ &\ & \quad + \,
A\,{|A'|}\sin(\phi_{A'})
\Big[
C^2 {|B'|}^2 + B^2{C'}^2 - 2\,C\,B\,{C'}\,{|B'|}\cos(\phi_{B'})
\Big]
\bigg\}\,.
\label{jv}
\end{eqnarray}

The mass matrices $M_{\rm u}$ and $M_{\rm d}$ in Eq.~(\ref{13}) do not
have positive definite eigenvalues. However, a viable mass matrix does
not need to have positive definite eigenvalues~\cite{fritzsch78}. These
eigenvalues are real but not necessarily positive. Thus, $\lambda^2_{\rm
u}=m^2_{\rm u}$, $\lambda^2_{\rm d}=m^2_{\rm d}$, etc., where $m_{\rm u}$
is the (positive) mass of the up quark, etc.

In this case it is possible to fix the relative phases between the
eigenvalues and the quark masses. For the up quark sector (and
analogously for the down one) we need a solution with the mass hierarchy
$|\lambda_{\rm u}|<<|\lambda_{\rm c}|<<|\lambda_{\rm t}|$. From the first
relation~(\ref{15}) we see that $\lambda_{\rm t}=m_{\rm t}$, $C$ being
positive. This coupled with the second relation in~(\ref{15}) require
$\lambda_{\rm u}=m_{\rm u}$ and $\lambda_{\rm c}=-m_{\rm c}<0$. Then, for
the Fritzsch case the relative phases between $\lambda$'s and $m$'s are

\begin{equation}
(\lambda_{\rm u}\,,\lambda_{\rm c}\,,\lambda_{\rm t}) =
(m_{\rm u}\,,-m_{\rm c}\,,m_{\rm t})
\quad
{\rm and}
\quad
(\lambda_{\rm d}\,,\lambda_{\rm s}\,,\lambda_{\rm b}) =
(m_{\rm d}\,,-m_{\rm s}\,,m_{\rm b})\,.
\label{17p}
\end{equation}

From the above formulation, we are now in position to apply the procedure
described at the end of Sec.~\ref{introduction} to determine the
parameters of the quark mass matrices in the Fritzsch case. The
parameters to be estimated are the six eigenvalues ($\lambda_{\rm
u}$,$\lambda_{\rm c}$,$\lambda_{\rm t}$) and ($\lambda_{\rm
d}$,$\lambda_{\rm s}$,$\lambda_{\rm b}$), and the two phases $\phi_{A'}$
and $\phi_{B'}$.

Using Eqs.~(\ref{15b}) for the up and down quark sectors the
$|V_{\alpha\, j}|$ and $J(V)$ can be expressed as functions of the six
eigenvalues and the phases $\phi_{A'}$ and $\phi_{B'}$. We now fit these
theoretical expressions to the experimental values of the four moduli
$|V_{\rm ud}|$, $|V_{\rm us}|$, $|V_{\rm cd}|$, and $|V_{\rm cs}|$, and
$J(V)$~\cite{pdg06}. In doing so we constrain the eigenvalues to the experimentally
determined values of the quark masses at the $M_Z$
energy scale~\cite{xing07} displayed in Column~2 of
Table~\ref{table1}, with the relative phases as given
by~(\ref{17p}). The fitted values obtained for the eigenvalues
are given in Column~3 of Table~\ref{table1}. Column~3 also gives
the values predicted for $\phi_{A'}$, $\phi_{B'}$, the four
moduli and $J(V)$. The corresponding $\Delta\chi^2$ for the
eigenvalues, $|V_{\alpha\, j}|$, and $J(V)$ are given in the
last column. Note that $\phi_{A'}$ and $\phi_{B'}$ are unknown
to begin with. The total $\chi^2/({\rm dof}) = 4.23/3=1.41$.

Also, from the relations in Eq.~(\ref{15b}) for the up and down quark
sectors and the entries of Table~\ref{table1} we can determine for the
derived parameters $A$, $B$, $C$, $|A'|$, $|B'|$, and $C'$, their \lq\lq
experimental" values (using the experimental constraints on the quark
masses) and their predicted values (using the fitted values of the
eigenvalues), along with their corresponding $\Delta\chi^2$
contributions. These numbers are shown in Table~\ref{table1b}.

We observe from Table~\ref{table1} that the fitted values obtained for
the phases $\phi_{A'}$ and $\phi_{B'}$ are compatible with
$-\pi/2$ and $0$, respectively. In Tables~\ref{table1c} and
\ref{table1d} we display the corresponding results obtained for this
particular choice. In this six parameters fit (the six eigenvalues) the
total $\chi^2/({\rm dof}) = 4.84/5=0.97$.

\section{\label{stech}Stech type mass matrix}

The second case we consider is the model of Stech~\cite{stech83},

\begin{equation}
M_{\rm u}=
\begin{pmatrix}
\lambda_{\rm u} & 0 & 0 \\
0 & \lambda_{\rm c} & 0 \\
0 & 0 & \lambda_{\rm t}
\end{pmatrix}
,
\quad\quad
M_{\rm d}=p\,M_{\rm u}+i\,
\begin{pmatrix}
0 & a & d \\
-a & 0 & b \\
-d & -b & 0
\end{pmatrix}
.
\label{13st}
\end{equation}

\noindent
The mass matrices are hermitian, the $\lambda$'s are the
eigenvalues of $M_{\rm u}$, and $p$ is constant. $a$, $b$, and $d$ are
real and $a$ and $b$ can be made positive like $A$ and $B$ in
Eq.~(\ref{13}). The Stech model has seven parameters.

Using the characteristic equation of $M_{\rm d}$, we may express $p$, $a$
and $b$ in terms of the eigenvalues of $M_{\rm u}$ and $M_{\rm d}$ and
the parameter $d$,

\begin{equation}
p=\frac{\lambda_{\rm d} + \lambda_{\rm s} + \lambda_{\rm b}}
{\lambda_{\rm u} + \lambda_{\rm c} + \lambda_{\rm t}},
\label{17st}
\end{equation}

\begin{equation}
a^2=\frac{-\lambda_{\rm u}\,E_1 + E_2 - (\lambda_{\rm c}-\lambda_{\rm
u})\,d^2}{\lambda_{\rm t}-\lambda_{\rm u}},
\label{14st}
\end{equation}

\begin{equation}
b^2=\frac{\lambda_{\rm t}\,E_1 - E_2 - (\lambda_{\rm t}-\lambda_{\rm
c})\,d^2}{\lambda_{\rm t}-\lambda_{\rm u}},
\label{15st}
\end{equation}

\noindent
where,

\begin{equation}
E_1=\frac{1}{2}[\lambda^2_{\rm d} + \lambda^2_{\rm s} + \lambda^2_{\rm b}
-p^2(\lambda^2_{\rm u} + \lambda^2_{\rm c} + \lambda^2_{\rm t})],
\quad
E_2=\frac{1}{p}(p^3\lambda_{\rm u}\,\lambda_{\rm c}\,\lambda_{\rm
t} - \lambda_{\rm d}\,\lambda_{\rm s}\,\lambda_{\rm b}).
\label{16st}
\end{equation}

From~(\ref{5}), the moduli of the unitary quark mixing matrix elements in
this case are,

\begin{eqnarray}
|V_{\alpha\,j}|^2&=&
\big{[}
(\lambda_k - \lambda_j)
(\lambda_l - \lambda_j)\big{]}^{-1}\times
\nonumber
\\
&\ &
\big{[}
(\lambda_k - p\,\lambda_{\alpha})(\lambda_l - p\,\lambda_{\alpha}) +
(a^2 + d^2)\,\delta_{\alpha,{\rm u}}
+
(a^2 + b^2)\,\delta_{\alpha,{\rm c}}
+
(b^2 + d^2)\,\delta_{\alpha,{\rm t}}
\big{]}
\,,
\label{vajst}
\end{eqnarray}

\noindent
again ($j$,$k$,$l$) is any permutation of (d,s,b) and we shall use
only the four independent expressions corresponding to the best
experimentally measured magnitudes $|V_{\rm ud}|$, $|V_{\rm us}|$,
$|V_{\rm cd}|$, and $|V_{\rm cs}|$\,.

For the Stech case the Jarslkog invariant $J(V)$ given by (\ref{12}) is
simply~\cite{chaturvedi06}

\begin{equation}
J(V) =
\frac{a\,b\,d}{(\lambda_{\rm b} - \lambda_{\rm s}) (\lambda_{\rm b} -
\lambda_{\rm d})(\lambda_{\rm s} - \lambda_{\rm d})}.
\label{jvst}
\end{equation}

In order to determine the parameters of the mass matrices in the Stech
case, notice first that using Eqs.~(\ref{17st}) - (\ref{16st}) the
original set of seven parameters ($\lambda_{\rm u}$, $\lambda_{\rm c}$,
$\lambda_{\rm t}$, $p$, $a$, $b$, $d$) can be replaced by the set
($\lambda_{\rm u}$, $\lambda_{\rm c}$, $\lambda_{\rm t}$, $\lambda_{\rm
d}$, $\lambda_{\rm s}$, $\lambda_{\rm b}$, $d$).

As mentioned before, $\lambda^2_{\rm u}=m^2_{\rm u}$, $\lambda^2_{\rm
d}=m^2_{\rm d}$, etc., and we need a solution with the mass hierarchies
$|\lambda_{\rm u}|<<|\lambda_{\rm c}|<<|\lambda_{\rm t}|$ for the up
quark sector and $|\lambda_{\rm d}|<<|\lambda_{\rm s}|<<|\lambda_{\rm
b}|$ for the down one. In this case the relative phases between the
$\lambda$'s and the quark masses can not be fixed {\it a priori} and all
the two possible signs in front of each of the quark masses must be
explored.

As in Sec.~\ref{fritzsch}, in the $\chi^2$ function we shall compare the
theoretical expressions of $|V_{\rm ud}|$\,, $|V_{\rm us}|$\,, $|V_{\rm
cd}|$\,, $|V_{\rm cs}|$ (from~(\ref{vajst})), and $J(V)$
(from~(\ref{jvst})) as functions of the six eigenvalues and $d$ (using
(\ref{17st}) - (\ref{16st})) with their experimental
counterparts~\cite{pdg06}. The $|\lambda|$'s are constrained to the
experimentally determined quark masses~\cite{xing07}.

Our best fit for the Stech case is obtained for the
identification~(\ref{17p}) of the relative phases between $\lambda$'s
and $m$'s, that is,

\begin{equation}
(\lambda_{\rm u}\,,\lambda_{\rm c}\,,\lambda_{\rm t}) =
(m_{\rm u}\,,-m_{\rm c}\,,m_{\rm t})\,,
\quad
(\lambda_{\rm d}\,,\lambda_{\rm s}\,,\lambda_{\rm b}) =
(m_{\rm d}\,,-m_{\rm s}\,,m_{\rm b})\,.
\label{17pst}
\end{equation}

\noindent
The corresponding numerical results are displayed in
Table~\ref{table2}. The total $\chi^2/({\rm dof}) = 9.10/4=2.28$.
Using the experimental constraints and the values of the fitted
parameters of Table~\ref{table2}, in Table~\ref{table2b} we show
the experimental, predicted, and $\Delta\chi^2$ values for the
derived parameter $p$ of Eq.~(\ref{17st}), and the predicted
values for $a$ and $b$ of Eqs.~(\ref{14st}) and (\ref{15st}),
respectively.

\section{\label{diagonal}new type os mass matrices}

Recently~\cite{chaturvedi07} mass matrices which are a simple
generalization of the Fritzsch mass matrix with one extra parameter in
$M_{\rm u}$ and $M_{\rm d}$ were considered. For want of a short name we
call it the CGS type mass matrix. The extra parameter comes from choosing
the 13 (and 31) matrix element to be non-zero and complex. This choice
gives CP-violation (see Eq.~(\ref{jdiag})), unlike the Fritzsch case,
even if one works in either up-quark or down-quark diagonal basis. In
subsections~\ref{down} and \ref{up} we consider this type of matrix in
the down-quark and up-quark diagonal basis, respectively. In
Sec.~\ref{nondiagonal} we do present fits with the CGS-type matrices in
the physical basis.

We now consider a basis in which the up-quark (down-quark) mass matrix
$M_{\rm u}$ ($M_{\rm d}$) is diagonal and the remaining mass matrix
$M_{\rm d}$ ($M_{\rm u}$) is hermitian of the CGS-type, namely,

\begin{equation}
M=
\begin{pmatrix}
0 & a & d \\
a^* & 0 & b \\
d^* & b^* & c
\end{pmatrix}
.
\label{di13}
\end{equation}

\noindent
For $d=0$ this reduces to the Fritzsch-type mass matrix and will give
$J(U)=0$, where $U$ is the corresponding diagonalizing unitary matrix.

Let $\lambda_{1,2,3}$ the eigenvalues of $M$, from the characteristic
equation we have

\[
c=\lambda_1+\lambda_2+\lambda_3\,,
\]

\begin{equation}
-(|a|^2+|b|^2+|d|^2)=\lambda_1\lambda_2+
\lambda_1\lambda_3+\lambda_2\lambda_3\,,
\label{di15}
\end{equation}

\[
-c|a|^2+2{\rm Re}(abd^*)=\lambda_1\lambda_2\lambda_3\,.
\]

\noindent As before, for the quark sector we need the mass hierarchy
$|\lambda_1|<<|\lambda_2|<<|\lambda_3|$ and from Eqs.~(\ref{di15}) it is
required that $\lambda_1,\lambda_3>0$ and $\lambda_2<0$, assuming $c>0$,
for both up and down quarks. For simplicity we take $a$ and $b$ to be
real and positive and $d$ as pure imaginary. Solving for $a$, $b$, and
$c$ yields

\begin{equation}
a=\left[
-
\frac{\lambda_1\lambda_2\lambda_3}
{\lambda_1+\lambda_2+\lambda_3}
\right]^{1/2}
\,,
\qquad
c=\lambda_1+\lambda_2+\lambda_3
\,,
\label{di15b}
\end{equation}

\begin{equation}
b=\left[-
\frac{
(\lambda_{\rm 3}+\lambda_{\rm 2})
(\lambda_{\rm 3}+\lambda_{\rm 1})
(\lambda_{\rm 2}+\lambda_{\rm 1})}
{\lambda_{\rm 1}+\lambda_{\rm 2}+\lambda_{\rm 3}
} - |d|^2
\right]^{1/2}
\,.
\label{di15c}
\end{equation}

\noindent
The CGS-type mass matrix in the up or down quark diagonal basis has in
principle, only 4 parameters, the eigenvalues $\lambda_1$,
$\lambda_2$, and $\lambda_3$ of $M$, and the magnitude of  $d$.
Nevertheless, the total number of parameters is 7 because this
number comes from the non-zero elements of the two mass matrices. These
are 3 parameters from the diagonal mass matrix plus the 4 in the
other non-diagonal mass matrix. The number of constraints is 11 as
before. We now investigate the viability of $M$ in both the up-quark
and down-quark diagonal basis.

\subsection{\label{down}CGS-type mass matrix in down quark diagonal
basis}

In this case $M=M_{\rm u}$ is the up-quark mass matrix which is
diagonalized by $V_{\rm u}$. So the CKM-matrix $V=V_{\rm u}$
since $V_{\rm d}=I$. According to~(\ref{5}), the magnitudes of the
unitary quark mixing matrix elements are given by

\begin{eqnarray}
|V_{\alpha\,j}|^2&=&
\big{[}
(\lambda_{\beta} - \lambda_{\alpha})(\lambda_{\gamma} - \lambda_{\alpha})
\big{]}^{-1}\times
\nonumber
\\
&\ &
\big{\{}
(a^2 + |d|^2 + \lambda_{\beta}\lambda_{\gamma})\,\delta_{j,{\rm d}}
+
(a^2 + b^2 + \lambda_{\beta}\lambda_{\gamma})\,\delta_{j,{\rm s}}
\label{vajdown}
\\
&\ & +
\big{[}b^2 + |d|^2 + (\lambda_{\beta} + \lambda_{\alpha})(\lambda_{\gamma} +
\lambda_{\alpha})\big{]}\,\delta_{j,{\rm b}}
\big{\}}
\,,
\nonumber
\end{eqnarray}

\noindent
where ($\alpha$,$\beta$,$\gamma$) is any permutation of (u,c,t). As
mentioned in Sec.~\ref{introduction}, we shall use the four best
experimentally measured magnitudes $|V_{\rm ud}|$, $|V_{\rm us}|$,
$|V_{\rm cd}|$, and $|V_{\rm cs}|$\,.

Note that $\lambda_{\rm c}<0$ would imply $J(V)=J(V_{\rm u})<0$, so we
choose $d=-i|d|$ in this case. The Jarslkog invariant $J(V)$ given by
(\ref{12}), is then~\cite{chaturvedi06}

\begin{equation}
J(V) = -
\frac{a\,b\,|d|}{(\lambda_{\rm t} - \lambda_{\rm c}) (\lambda_{\rm t} -
\lambda_{\rm u})(\lambda_{\rm c} - \lambda_{\rm u})}.
\label{jvdown}
\end{equation}

The relative phases between the eigenvalues and the up quark masses are

\begin{equation}
(\lambda_{\rm u}\,,\lambda_{\rm c}\,,\lambda_{\rm t}) =
(m_{\rm u}\,,-m_{\rm c}\,,m_{\rm t})\,.
\label{17pdown}
\end{equation}

\noindent
The parameters to be estimated in this case are the three
eigenvalues ($\lambda_{\rm u}$,$\lambda_{\rm c}$,$\lambda_{\rm t}$) and
$|d|$.

In the $\chi^2$ function we shall compare the theoretical expressions of
$|V_{\rm ud}|$\,, $|V_{\rm us}|$\,, $|V_{\rm cd}|$\,, $|V_{\rm cs}|$
(from (\ref{vajdown})), and $J(V)$ (from (\ref{jvdown})) as functions of
the above parameters (using relations (\ref{di15b}) and (\ref{di15c})),
with their experimental counterparts~\cite{pdg06}. We shall also
constrain the $\lambda$'s to the experimentally determined quark
masses~\cite{xing07} with the relative phases as given
by~(\ref{17pdown}).

The results of our best fit with the four parameters and the eight
constraints used are displayed in Table~\ref{table3}. The total
$\chi^2/({\rm dof}) = 5.92/4=1.48$. Using the experimental
constraints and the values of the fitted parameters of
Table~\ref{table3}, in Table~\ref{table3b} we show the
experimental, predicted, and $\Delta\chi^2$ values for the
derived parameters $a$ and $c$ of Eqs.~(\ref{di15b}) and the
predicted value for $b$ from Eq.~(\ref{di15c}).

\subsection{\label{up}CGS-type mass matrix in up quark diagonal basis}

In this case $M=M_{\rm d}$ is the down-quark mass matrix which is
diagonalized by $V_{\rm d}$. So the CKM-matrix $V=V^{\dag}_{\rm d}$
since $V_{\rm u}=I$. According to~(\ref{5}), the magnitudes of the
unitary quark mixing matrix elements are given by

\begin{eqnarray}
|V_{\alpha\,j}|^2&=&
\big{[}
(\lambda_{k} - \lambda_{j})(\lambda_{l} - \lambda_{j})
\big{]}^{-1}\times
\nonumber
\\
&\ &
\big{\{}
(a^2 + |d|^2 + \lambda_{k}\lambda_{l})\,\delta_{\alpha,{\rm u}}
+
(a^2 + b^2 + \lambda_{k}\lambda_{l})\,\delta_{\alpha,{\rm c}}
\label{vajup}
\\
&\ & +
\big{[}b^2 + |d|^2 + (\lambda_{k} + \lambda_{j})(\lambda_{l} +
\lambda_{j})\big{]}\,\delta_{\alpha,{\rm t}}
\big{\}}
\,,
\nonumber
\end{eqnarray}

\noindent
where ($j$,$k$,$l$) is any permutation of (d,s,b). As
mentioned in Sec.~\ref{introduction}, we shall use the four best
experimentally measured magnitudes $|V_{\rm ud}|$, $|V_{\rm us}|$,
$|V_{\rm cd}|$, and $|V_{\rm cs}|$\,.

Note that $\lambda_{\rm s}<0$ imply $J(V)=J(V^{\dagger}_{\rm d})>0$, so
we choose $d=i|d|$ in this case. The Jarslkog invariant $J(V)$ given by
(\ref{12}), is then~\cite{chaturvedi06}

\begin{equation}
J(V) = -
\frac{
a\,b\,|d|
}{
(\lambda_{\rm b} - \lambda_{\rm s}) (\lambda_{\rm b} -
\lambda_{\rm d})(\lambda_{\rm s} - \lambda_{\rm d})
}.
\label{jvup}
\end{equation}

The relative phases between the eigenvalues and the down quark masses are

\begin{equation}
(\lambda_{\rm d}\,,\lambda_{\rm s}\,,\lambda_{\rm b}) =
(m_{\rm d}\,,-m_{\rm s}\,,m_{\rm b})\,.
\label{17pup}
\end{equation}

\noindent
The parameters to be estimated in this case are the three
eigenvalues ($\lambda_{\rm d}$,$\lambda_{\rm s}$,$\lambda_{\rm b}$) and
$|d|$.

In the $\chi^2$ function we shall compare the theoretical expressions of
$|V_{\rm ud}|$\,, $|V_{\rm us}|$\,, $|V_{\rm cd}|$\,, $|V_{\rm cs}|$
(from (\ref{vajup})), and $J(V)$ (from (\ref{jvup})) as functions of the
above parameters (using (\ref{di15b}) and (\ref{di15c})), with
their experimental counterparts~\cite{pdg06}. We shall also constrain
the $\lambda$'s to the experimentally determined quark
masses~\cite{xing07} with the relative phases as given
by~(\ref{17pup}).

The results of our best fit with the four parameters and the eight
constraints used are displayed in Table~\ref{table4}. The total
$\chi^2/({\rm dof}) = 15.50/4=3.88$. Using the experimental
constraints and the values of the fitted parameters of
Table~\ref{table4}, in Table~\ref{table4b} we show the
experimental, predicted, and $\Delta\chi^2$ values for the
derived parameters $a$ and $c$ of Eqs.~(\ref{di15b}) and the
predicted value for $b$ from Eq.~(\ref{di15c}).

\section{\label{nondiagonal}CGS-type mass matrix in physical basis}

We conclude by considering a small variation of the Fritzsch case, given
by the hermitian matrices

\begin{equation}
M_{\rm u}=
\begin{pmatrix}
0 & a & 0 \\
a & 0 & b \\
0 & b & c
\end{pmatrix}
,
\quad\quad
M_{\rm d}=
\begin{pmatrix}
0 & a' & i|d'| \\
a' & 0 & b' \\
-i|d'| & b' & c'
\end{pmatrix}
.
\label{13non}
\end{equation}

\noindent
The parameters $a$, $b$, $c$, $a'$, $b'$, and $c'$ are considered real
and positive. Here $M_{\rm d}$ is CGS-type and $M_{\rm u}$ is
Fritzsch-type so that we have 7 real parameters.

From the characteristic equations of $M_{\rm u}$ and $M_{\rm d}$ we can
solve for the matrix elements in terms of the corresponding eigenvalues
and $|d'|$,

\[
a=\left[
-\frac{\lambda_{\rm u}\lambda_{\rm c}\lambda_{\rm t}}
{\lambda_{\rm u}+\lambda_{\rm c}+\lambda_{\rm t}
}
\right]^{1/2}
,
\qquad
c=\lambda_{\rm u}+\lambda_{\rm c}+\lambda_{\rm t}\,,
\]

\begin{equation}
b=\left[-
\frac{
(\lambda_{\rm t}+\lambda_{\rm c})
(\lambda_{\rm t}+\lambda_{\rm u})
(\lambda_{\rm c}+\lambda_{\rm u})}
{\lambda_{\rm u}+\lambda_{\rm c}+\lambda_{\rm t}
}
\right]^{1/2}
,
\label{15bnon}
\end{equation}

\noindent
and

\[
a'=\left[
-\frac{\lambda_{\rm d}\lambda_{\rm s}\lambda_{\rm b}}
{\lambda_{\rm d}+\lambda_{\rm s}+\lambda_{\rm b}
}
\right]^{1/2}
,
\qquad
c'=\lambda_{\rm d}+\lambda_{\rm s}+\lambda_{\rm b}\,,
\]

\begin{equation}
b'=\left[-
\frac{
(\lambda_{\rm b}+\lambda_{\rm s})
(\lambda_{\rm b}+\lambda_{\rm d})
(\lambda_{\rm s}+\lambda_{\rm d})}
{\lambda_{\rm d}+\lambda_{\rm s}+\lambda_{\rm b}
} - |d'|^2
\right]^{1/2}
.
\label{15bpnon}
\end{equation}

From~(\ref{5}), the magnitudes of the unitary quark
mixing matrix elements in this case are

\begin{eqnarray}
|V_{\alpha\,j}|^2&=&
\big{[}(\lambda_{\beta} - \lambda_{\alpha})
(\lambda_{\gamma} - \lambda_{\alpha})
(\lambda_k - \lambda_j)
(\lambda_l - \lambda_j)\big{]}^{-1}\times
\nonumber
\\
&\ &
\big{[}
(a\,b'+b\,a')^2
+ 2(a^2a'^2+b^2b'^2)+(a^2+b^2)|d'|^2
\nonumber
\\
&\ &+\,
2\,a\,a'(\lambda_{\beta} + \lambda_{\gamma})
(\lambda_k + \lambda_l)
+\lambda_{\beta}\lambda_{\gamma}
\lambda_{k} \lambda_l
\nonumber
\\
&\ &+\,
\lambda_{k} \lambda_l
(
a^2+b^2+\lambda^2_{\alpha}
)+
\lambda_{\beta}\lambda_{\gamma}
(
a'^2+b'^2+|d'|^2+\lambda^2_j
)
\nonumber
\\
&\ &+\,
c\,\lambda_{\alpha}(b'^2+|d'|^2)
+c'\lambda_{j}\,b^2
+\lambda_{\alpha}\lambda_{j}(c\,c'+2b\,b')
\big{]}
\,,
\label{vajnon}
\end{eqnarray}

\noindent
where ($\alpha$,$\beta$,$\gamma$) is any permutation of (u,c,t)
and ($j$,$k$,$l$) any permutation of (d,s,b). We shall use the four best
experimentally measured magnitudes $|V_{\rm ud}|$, $|V_{\rm us}|$,
$|V_{\rm cd}|$, and $|V_{\rm cs}|$\,.

The Jarslkog invariant $J(V)$ given by Eq.~(\ref{12}) translates
into,

\begin{equation}
J(V) = -
\frac{
b\,|d'|\,\big{[}(a'b - a\,b')\,(b\,c' - b'c) +
a\,c\,|d'|^2\big{]}
}{
(\lambda_{\rm b} - \lambda_{\rm s}) (\lambda_{\rm b} -
\lambda_{\rm d})(\lambda_{\rm s} - \lambda_{\rm d})
(\lambda_{\rm t} - \lambda_{\rm c}) (\lambda_{\rm t} -
\lambda_{\rm u})(\lambda_{\rm c} - \lambda_{\rm u})
}.
\label{jvnon}
\end{equation}

As in the Fritzsch case it is possible to fix the relative phases between
the eigenvalues and the quark masses using Eqs.~(\ref{15bnon}) and
(\ref{15bpnon}),

\begin{equation}
(\lambda_{\rm u}\,,\lambda_{\rm c}\,,\lambda_{\rm t}) =
(m_{\rm u}\,,-m_{\rm c}\,,m_{\rm t})
\quad
{\rm and}
\quad
(\lambda_{\rm d}\,,\lambda_{\rm s}\,,\lambda_{\rm b}) =
(m_{\rm d}\,,-m_{\rm s}\,,m_{\rm b})\,.
\label{17pnon}
\end{equation}

\noindent
The parameters to be estimated are the six eigenvalues ($\lambda_{\rm
u}$,$\lambda_{\rm c}$,$\lambda_{\rm t}$) and ($\lambda_{\rm
d}$,$\lambda_{\rm s}$,$\lambda_{\rm b}$), and $|d'|$.

In the $\chi^2$ function we shall compare the theoretical expressions of
$|V_{\rm ud}|$\,, $|V_{\rm us}|$\,, $|V_{\rm cd}|$\,, $|V_{\rm cs}|$
(from (\ref{vajnon})), and $J(V)$ (from (\ref{jvnon})) as functions of
the above parameters (using relations (\ref{15bnon}) and
(\ref{15bpnon})), with their experimental counterparts~\cite{pdg06}.
We shall also constrain the $\lambda$'s to the experimentally
determined quark masses~\cite{xing07} with the relative phases as
given by~(\ref{17pnon}).

The results of our best fit are displayed in Table~\ref{table5}.
Using the experimental constraints and the values of the fitted
parameters of Table~\ref{table5}, in Table~\ref{table5b} we show
the experimental, predicted, and $\Delta\chi^2$ values for the
derived parameters $a$, $b$, $c$, $a'$, and $c'$ of
Eqs.~(\ref{15bnon}) and (\ref{15bpnon}), and the predicted value
for $b'$ of Eq.~(\ref{15bpnon}).

For this case, the total $\chi^2/({\rm dof}) = 1.89/4=0.47$ which
is much better than the Fritzsch case or the Stech case. The fit
for the mixed case, namely $M_{\rm u}$ CGS-type and $M_{\rm d}$
Fritzsch-type (with 7 real parameters) gives the same
$\chi^2/({\rm dof})$ as the fit given above. The reason is that
there exists an unitary matrix $Y$ such that it can rotate the
mass matrices in Eq.~(\ref{13non}) to mass matrices $N_{\rm
u}=Y^{\dagger}M_{\rm u}Y$ and $N_{\rm d}=Y^{\dagger}M_{\rm d}Y$
which are CGS-type and Fritzsch-type, respectively. Similarly,
we can rotate $M_{\rm u}$ and $M_{\rm d}$ in Eq.~(\ref{13non}),
using an unitary matrix so that both the up and down quark mass
matrices are of the CGS-type. Consequently, the case when both
mass matrices are of the CGS-type with 8 parameters does not
give any improvement in the quality of the fit.

\section{\label{conclusions}Concluding Remarks and discussion of results}
Working with different specific textures of quark mass matrices, in this
work we have determined bounds on the mass matrices elements for each
case using as experimental restrictions the values of the four
independent best measured moduli of the quark mixing matrix
elements, the Jarlskog invariant $J(V)$, and the quark masses at
the $M_Z$ energy scale.

The results of the fits are presented in
Tables~\ref{table1}-\ref{table5b}. The  $\chi^2/({\rm dof})$ for the
various cases are tabulated in Table~\ref{table6} for
easy comparison. As can be seen, the fits for the Fritzsch and
Stech type are comparable even though the former has 8
parameters compared to the 7 of the latter. The CGS-type
matrix in the $M_{\rm u}$ diagonal basis is poorer despite having
7 parameters, but in the $M_{\rm d}$ diagonal basis it gives a
fit comparable to the Fritzsch and Stech type matrices. However,
the choice of different type of mass matrices in
Sec.~\ref{nondiagonal} (last row of Table~\ref{table6}) for up
and down quarks gives a very good fit despite having only 7
parameters. This is very encouraging phenomenologically.

An important question is whether the bounds obtained on the
structure of the quark mass matrices are affected by the
evolution of the quark masses. To check the stability of our type
of analysis we repeated it using as restrictions the values
of the quark masses at the 2~GeV energy scale~\cite{xing07}. The
results for $\chi^2/({\rm dof})$ for the various cases, at 2~GeV
scale, are summarized in Table~\ref{table6-2gev}. Results in
Table~\ref{table6-2gev} are very similar to the results in
Table~\ref{table6} implying that our results are not affected by
the evolution of the quark masses from 2 to
91~GeV~\cite{footnote}.

A simply way of understanding this behaviour is to note that if
all quark masses are scaled by a common factor, then the
algebraic expressions for the dimensionless numbers $J(V)$ and
the four moduli $|V_{\alpha\, j}|$ ($\alpha={\rm u}, {\rm c}$;
$j={\rm d}, {\rm s}$) will be unaffected. Actually, the ratio of
the quark masses at 2~GeV and $M_Z$ energy scales are $m_q(2~{\rm
GeV})/m_q(M_Z) = 1.71 - 1.74$ ($q = {\rm u}, {\rm d},
{\rm s}, {\rm c}, {\rm b}$) and 1.85 for  $q = {\rm
t}$~\cite{xing07}.

Phenomenologically, our results are encouraging. However, the
basic problem remains, namely to find a fundamental theoretical
mechanism to generate quark mass matrices.

\begin{acknowledgments}

One of us (G.~S-C) is grateful to the Faculty of Mathematics,
Autonomous University of Yucat\'an, M\'exico, for hospitality where part
of this work was done. VG and G.~S-C would like to thank CONACyT
(M\'exico) for partial support.

\end{acknowledgments}

\clearpage

\begin{table}

\caption{ Fritzsch-type mass matrices in physical basis. In the first
part of the table we show the experimental constraints imposed (where
applicable) and the values obtained for the fitted parameters, along
with their $\Delta\chi^2$ contribution. The eigenvalues $\lambda$ are
in MeV. The experimentally observed, predicted, and $\Delta\chi^2$
values of the four best measured magnitudes of the quark mixing
matrix elements and the Jarlskog invariant are displayed in the
second part of the table. The total $\chi^2/({\rm dof}) =
4.23/3=1.41$. \label{table1} }

\begin{ruledtabular}

\begin{tabular}{c|c|c|c}

Fitted & Experimental & Value &
$\Delta\chi^2$ \\

parameters & constraint~(\ref{17p}),\cite{xing07} &  obtained
from fit & \\

\hline

$\lambda_{\rm u}$ & $1.28^{+0.50}_{-0.43}$ &
$1.66^{+0.40}_{-0.36}$ & $0.58$ \\

$\lambda_{\rm c}$ & $-(0.624\pm 0.083)\times
10^{3}$ & $-(0.621\pm 0.080)\times
10^{3}$ & 0.0013 \\

$\lambda_{\rm t}$ & $(172.5\pm 3.0)\times
10^{3}$ & $(172.4\pm 3.0)\times
10^{3}$ & 0.0011 \\

$\lambda_{\rm d}$ & $2.91^{+1.24}_{-1.20}$ &
$2.29^{+0.50}_{-0.49}$ & $0.27$ \\

$\lambda_{\rm s}$ & $-(55^{+16}_{-15})$ &
$-(38.0^{+4.3}_{-5.8})$ & $1.28$ \\

$\lambda_{\rm b}$ & $(2.89\pm 0.09)\times
10^{3}$ & $(2.901\pm 0.090)\times
10^{3}$ & 0.015 \\

$\phi_{A'}$ & --- & $(-71^{+19}_{-23})^\circ$ & --- \\

$\phi_{B'}$ & --- & $(-4^{+18}_{-16})^\circ$ & --- \\

\hline

Observable & Experiment~\cite{pdg06} &
Prediction~(\ref{vaj}),(\ref{jv}) & $\Delta\chi^2$ \\

\hline

$|V_{\rm ud}|$ & $0.97383\pm
0.00024$ & $0.97391$ & $0.11$ \\

$|V_{\rm us}|$ & $0.2272\pm
0.0010$ & $0.2269$ & $0.090$ \\

$|V_{\rm cd}|$ & $0.2271\pm
0.0010$ & $0.2266$ & $0.25$ \\

$|V_{\rm cs}|$ & $0.97296\pm
0.00024$ & $0.97266$ & $1.56$ \\

$J(V)$ & $(3.08\pm 0.18)\times 10^{-5}$ &
$3.03\times 10^{-5}$ & $0.077$ \\

\end{tabular}

\end{ruledtabular}

\end{table}

\begin{table}

\caption{ Fritzsch-type mass matrices in physical basis.
Experimentally observed, predicted, and $\Delta\chi^2$ values of the
derived parameters $A$, $B$, $C$, and $|A'|$, $|B'|$, $C'$, of
Eqs.~(\ref{15b}) in the up and down quark sectors, respectively.
Experimental and predicted values were determined by using the
entries of Table~\ref{table1} for the experimental quark masses
constraints and the fitted values of the eigenvalues, respectively.
\label{table1b} }

\begin{ruledtabular}

\begin{tabular}{c|c|c|c}

Derived & Experiment & Prediction & $\Delta\chi^2$ \\

parameters & (MeV) & (MeV) & \\

\hline

$A$ & $28.3^{+5.8}_{-5.1}$ & $32.1$ & 0.43\\

$B$ & $(10.36\pm 0.70)\times 10^{3}$ & $10.33\times 10^{3}$ &
0.0018 \\

$C$ & $(171.9\pm 3.0)\times 10^{3}$ & $171.8\times 10^{3}$ &
0.0011 \\

$|A'|$ & $12.8^{+3.3}_{-3.2}$ & $9.4$ & 1.13 \\

$|B'|$ & $388^{+60}_{-56}$ & $322$ & 1.39 \\

$C'$ & $(2.838\pm 0.091)\times 10^{3}$ & $2.865\times 10^{3}$ &
0.088 \\

\end{tabular}

\end{ruledtabular}

\end{table}

\begin{table}

\caption{ Fritzsch-type mass matrices in physical basis with the
phases fixed at $\phi_{A'}=-\pi/2$ and $\phi_{B'}=0$. In the first
part of the table we show the experimental constraints imposed (where
applicable) and the values obtained for the fitted parameters, along
with their $\Delta\chi^2$ contribution. The eigenvalues $\lambda$ are
in MeV. The experimentally observed, predicted, and $\Delta\chi^2$
values of the four best measured magnitudes of the quark mixing
matrix elements and the Jarlskog invariant are displayed in the
second part of the table. The total $\chi^2/({\rm dof}) =
4.84/5=0.97$.  \label{table1c} }

\begin{ruledtabular}

\begin{tabular}{c|c|c|c}

Fitted & Experimental & Value &
$\Delta\chi^2$ \\

parameters & constraint~(\ref{17p}),\cite{xing07} &  obtained
from fit & \\

\hline

$\lambda_{\rm u}$ & $1.28^{+0.50}_{-0.43}$ &
$1.66^{+0.40}_{-0.36}$ & $0.58$ \\

$\lambda_{\rm c}$ & $-(0.624\pm 0.083)\times
10^{3}$ & $-(0.622\pm 0.079)\times
10^{3}$ & 0.00058 \\

$\lambda_{\rm t}$ & $(172.5\pm 3.0)\times
10^{3}$ & $(172.4\pm 3.0)\times
10^{3}$ & 0.0011 \\

$\lambda_{\rm d}$ & $2.91^{+1.24}_{-1.20}$ &
$1.98\pm 0.22$ & $0.60$ \\

$\lambda_{\rm s}$ & $-(55^{+16}_{-15})$ &
$-(38.3^{+4.0}_{-3.9})$ & $1.24$ \\

$\lambda_{\rm b}$ & $(2.89\pm 0.09)\times
10^{3}$ & $(2.902\pm 0.090)\times
10^{3}$ & 0.018 \\

\hline

Observable & Experiment~\cite{pdg06} &
Prediction~(\ref{vaj}),(\ref{jv}) & $\Delta\chi^2$ \\

\hline

$|V_{\rm ud}|$ & $0.97383\pm
0.00024$ & $0.97393$ & $0.17$ \\

$|V_{\rm us}|$ & $0.2272\pm
0.0010$ & $0.2269$ & $0.090$ \\

$|V_{\rm cd}|$ & $0.2271\pm
0.0010$ & $0.2266$ & $0.25$ \\

$|V_{\rm cs}|$ & $0.97296\pm
0.00024$ & $0.97264$ & $1.78$ \\

$J(V)$ & $(3.08\pm 0.18)\times 10^{-5}$ &
$3.02\times 10^{-5}$ & $0.11$ \\

\end{tabular}

\end{ruledtabular}

\end{table}

\begin{table}

\caption{ Fritzsch-type mass matrices in physical basis with the phases
fixed at $\phi_{A'}=-\pi/2$ and $\phi_{B'}=0$. Experimentally
observed, predicted, and $\Delta\chi^2$ values of the derived parameters
$A$, $B$, $C$, and $|A'|$, $|B'|$, $C'$, of Eqs.~(\ref{15b}) in the up
and down quark sectors, respectively. Experimental and predicted values
were determined by using the entries of Table~\ref{table1} for the
experimental quark masses constraints and the fitted values of the
eigenvalues, respectively. \label{table1d} }

\begin{ruledtabular}

\begin{tabular}{c|c|c|c}

Derived & Experiment & Prediction & $\Delta\chi^2$ \\

parameters & (MeV) & (MeV) & \\

\hline

$A$ & $28.3^{+5.8}_{-5.1}$ & $32.2$ & 0.45 \\

$B$ & $(10.36\pm 0.70)\times 10^{3}$ & $10.34\times 10^{3}$ &
0.00082 \\

$C$ & $(171.9\pm 3.0)\times 10^{3}$ & $171.8\times 10^{3}$ &
0.0011 \\

$|A'|$ & $12.8^{+3.3}_{-3.2}$ & $8.8$ & 1.56 \\

$|B'|$ & $388^{+60}_{-56}$ & $325$ & 1.27 \\

$C'$ & $(2.838\pm 0.091)\times 10^{3}$ & $2.865\times 10^{3}$ &
0.088 \\

\end{tabular}

\end{ruledtabular}

\end{table}

\begin{table}

\caption{
Stech-type mass matrix in up quark diagonal basis. In the first part
of the table we show the experimental constraints imposed (where
applicable) and the values obtained for the fitted parameters, along
with their $\Delta\chi^2$ contribution. The eigenvalues $\lambda$
and $d$ are in MeV. The experimentally observed, predicted, and
$\Delta\chi^2$ values of the four best measured magnitudes of the
quark mixing matrix elements and the Jarlskog invariant are displayed
in the second part of the table. The total $\chi^2/({\rm dof}) =
9.10/4=2.28$.  \label{table2} }

\begin{ruledtabular}

\begin{tabular}{c|c|c|c}

\hline

Fitted & Experimental & Value & $\Delta\chi^2$
\\

parameters & constraint~(\ref{17p}),\cite{xing07} &  obtained
from fit & \\

\hline

$\lambda_{\rm u}$ & $1.28^{+0.50}_{-0.43}$ &
$1.29^{+0.23}_{-0.22}$ & $0.00040$ \\

$\lambda_{\rm c}$ & $-(0.624\pm 0.083)\times
10^{3}$ & $-(0.650^{+0.035}_{-0.034})\times
10^{3}$ & 0.10 \\

$\lambda_{\rm t}$ & $(172.5\pm 3.0)\times
10^{3}$ & $(172.4\pm 2.8)\times
10^{3}$ & 0.0011 \\

$\lambda_{\rm d}$ & $2.91^{+1.24}_{-1.20}$ &
$0.9570\pm 0.0045$ & $2.65$ \\

$\lambda_{\rm s}$ & $-(55^{+16}_{-15})$ &
$-(17.720^{+0.085}_{-0.086})$ & $6.18$ \\

$\lambda_{\rm b}$ & $(2.89\pm 0.09)\times
10^{3}$ & $(2.901^{+0.054}_{-0.052})\times
10^{3}$ & 0.015 \\

$d$ & --- & $-9.07\pm 0.43$ & --- \\

\hline

Observable & Experiment~\cite{pdg06} &
Prediction~(\ref{vajst}),(\ref{jvst}) & $\Delta\chi^2$ \\

\hline

$|V_{\rm ud}|$ & $0.97383\pm
0.00024$ & $0.97385$ & $0.0069$ \\

$|V_{\rm us}|$ & $0.2272\pm
0.0010$ & $0.2272$ & $0.00$ \\

$|V_{\rm cd}|$ & $0.2271\pm
0.0010$ & $0.2269$ & $0.040$ \\

$|V_{\rm cs}|$ & $0.97296\pm
0.00024$ & $0.97288$ & $0.11$ \\

$J(V)$ & $(3.08\pm 0.18)\times 10^{-5}$ &
$3.08\times 10^{-5}$ & $0.00$ \\

\end{tabular}

\end{ruledtabular}

\end{table}

\begin{table}

\caption{ Stech-type mass matrix in up quark diagonal basis.
Experimental, predicted, and $\Delta\chi^2$ values for the derived
parameter $p$ of Eq.~(\ref{17st}). The predicted values for $a$ and
$b$ of Eqs.~(\ref{14st}) and (\ref{15st}), respectively, are also
displayed. Experimental and predicted values were determined by using
the entries of Table~\ref{table2} for the experimental quark masses
constraints and the fitted values of the eigenvalues, respectively.
\label{table2b} }

\begin{ruledtabular}

\begin{tabular}{c|c|c|c}

Derived & Experiment & Prediction & ($\Delta\chi^2$) \\

parameters & & & \\

\hline

$p$ & $(16.51\pm 0.60)\times 10^{-3}$ & $16.79\times 10^{-3}$ &
0.22 \\

\hline

$a$ & --- & $4.12$\,MeV & --- \\

$b$ & --- & $130$\,MeV & --- \\

\end{tabular}

\end{ruledtabular}

\end{table}

\begin{table}

\caption{
CGS-type mass matrix in down quark diagonal basis. In the first part
of the table we show the experimental constraints imposed (where
applicable) and the values obtained for the fitted parameters, along
with their $\Delta\chi^2$ contribution. The eigenvalues $\lambda$
and $|d|$ are in MeV. The experimentally observed, predicted, and
$\Delta\chi^2$ values of the four best measured magnitudes of the
quark mixing matrix elements and the Jarlskog invariant are displayed
in the second part of the table.  The total $\chi^2/({\rm dof}) =
5.92/4=1.48$. \label{table3} }

\begin{ruledtabular}

\begin{tabular}{c|c|c|c}

\hline

Fitted & Experimental & Value & $\Delta\chi^2$
\\

parameters & constraint~(\ref{17p}),\cite{xing07} &  obtained
from fit & \\

\hline

$\lambda_{\rm u}$ & $1.28^{+0.50}_{-0.43}$ &
$1.26^{+0.24}_{-0.19}$ & $0.0022$ \\

$\lambda_{\rm c}$ & $-(0.624\pm 0.083)\times
10^{3}$ & $-(0.490^{+0.060}_{-0.059})\times
10^{3}$ & 2.61 \\

$\lambda_{\rm t}$ & $(172.5\pm 3.0)\times
10^{3}$ & $(173.0\pm 3.0)\times
10^{3}$ & 0.028 \\

$|d|$ & --- & $(2.04^{+0.13}_{-0.14})\times
10^{3}$ & --- \\

\hline

Observable & Experiment~\cite{pdg06} &
Prediction~(\ref{vajdown}),(\ref{jvdown}) & $\Delta\chi^2$ \\

\hline

$|V_{\rm ud}|$ & $0.97383\pm
0.00024$ & $0.97395$ & $0.25$ \\

$|V_{\rm us}|$ & $0.2272\pm
0.0010$ & $0.2268$ & $0.16$ \\

$|V_{\rm cd}|$ & $0.2271\pm
0.0010$ & $0.2265$ & $0.36$ \\

$|V_{\rm cs}|$ & $0.97296\pm
0.00024$ & $0.97258$ & $2.51$ \\

$J(V)$ & $(3.08\pm 0.18)\times 10^{-5}$ &
$3.08\times 10^{-5}$ & $0.00$ \\

\end{tabular}

\end{ruledtabular}

\end{table}

\begin{table}

\caption{ CGS-type mass matrix in down quark diagonal basis.
Experimental, predicted, and $\Delta\chi^2$ values for the derived
parameters $a$ and $c$, of Eqs.~(\ref{di15b}). The predicted value
for $b$ of Eq.~(\ref{di15c}) is also displayed. Experimental and
predicted values were determined by using the entries of
Table~\ref{table3} for the experimental quark masses constraints and
the fitted values of the eigenvalues, respectively. \label{table3b} }

\begin{ruledtabular}

\begin{tabular}{c|c|c|c}

Derived & Experiment & Prediction & ($\Delta\chi^2$) \\

parameters & (MeV) & (MeV) & \\

\hline

$a$ & $28.3^{+5.8}_{-5.1}$ & $24.9$ & 0.44 \\

$c$ & $(171.9\pm 3.0)\times 10^{3}$ & $172.5\times 10^{3}$ &
0.040 \\

\hline

$b$ & --- & $8.97\times 10^{3}$ & --- \\

\end{tabular}

\end{ruledtabular}

\end{table}

\begin{table}

\caption{ CGS-type mass matrix in up quark diagonal basis. In the
first part of the table we show the experimental constraints imposed
(where applicable) and the values obtained for the fitted parameters,
along with their $\Delta\chi^2$ contribution. The eigenvalues $\lambda$
and $|d|$ are in MeV. The experimentally observed, predicted, and
$\Delta\chi^2$ values of the four best measured magnitudes of the
quark mixing matrix elements and the Jarlskog invariant are displayed
in the second part of the table. The total $\chi^2/({\rm dof}) =
15.50/4=3.88$. \label{table4} }

\begin{ruledtabular}

\begin{tabular}{c|c|c|c}

\hline

Fitted & Experimental & Value & $\Delta\chi^2$
\\

parameters & constraint~(\ref{17p}),\cite{xing07} &  obtained
from fit & \\

\hline

$\lambda_{\rm d}$ & $2.91^{+1.24}_{-1.20}$ &
$0.318^{+0.099}_{-0.106}$ & $4.67$ \\

$\lambda_{\rm s}$ & $-(55^{+16}_{-15})$ &
$-(6.4\pm 1.7)$ & $10.50$ \\

$\lambda_{\rm b}$ & $(2.89\pm 0.09)\times
10^{3}$ & $(2.895\pm 0.090)\times
10^{3}$ & 0.0031 \\

$|d|$ & --- & $9.2^{+2.1}_{-1.3}$ & --- \\

\hline

Observable & Experiment~\cite{pdg06} &
Prediction~(\ref{vajup}),(\ref{jvup}) & $\Delta\chi^2$ \\

\hline

$|V_{\rm ud}|$ & $0.97383\pm
0.00024$ & $0.97387$ & $0.028$ \\

$|V_{\rm us}|$ & $0.2272\pm
0.0010$ & $0.2271$ & $0.010$ \\

$|V_{\rm cd}|$ & $0.2271\pm
0.0010$ & $0.2269$ & $0.040$ \\

$|V_{\rm cs}|$ & $0.97296\pm
0.00024$ & $0.97284$ & $0.25$ \\

$J(V)$ & $(3.08\pm 0.18)\times 10^{-5}$ &
$3.08\times 10^{-5}$ & $0.00$ \\

\end{tabular}

\end{ruledtabular}

\end{table}

\begin{table}

\caption{ CGS-type mass matrix in up quark diagonal basis.
Experimental, predicted, and $\Delta\chi^2$ values for the derived
parameters $a$ and $c$, of Eqs.~(\ref{di15b}). The predicted value
for $b$ of Eq.~(\ref{di15c}) is also displayed. Experimental and
predicted values were determined by using the entries of
Table~\ref{table4} for the experimental quark masses constraints and
the fitted values of the eigenvalues, respectively. \label{table4b} }

\begin{ruledtabular}

\begin{tabular}{c|c|c|c}

Derived & Experiment & Prediction & ($\Delta\chi^2$) \\

parameters & (MeV) & (MeV) & \\

\hline

$a$ & $12.8^{+3.3}_{-3.2}$ & $1.4$ & 12.69 \\

$c$ & $(2.838\pm 0.091)\times 10^{3}$ & $2.889\times 10^{3}$ &
0.31 \\

\hline

$b$ & --- & $133$ & --- \\

\end{tabular}

\end{ruledtabular}

\end{table}

\begin{table}

\caption{ Fritzsch- and CGS-type mass matrices in physical basis. In
the first part of the table we show the experimental constraints
imposed (where applicable) and the values obtained for the fitted
parameters, along with their $\Delta\chi^2$ contribution. The
eigenvalues $\lambda$ and $|d'|$ are in MeV. The experimentally
observed, predicted, and $\Delta\chi^2$ values of the four best
measured magnitudes of the quark mixing matrix elements and the
Jarlskog invariant are displayed in the second part of the table. The
total $\chi^2/({\rm dof}) = 1.89/4=0.47$.  \label{table5} }

\begin{ruledtabular}

\begin{tabular}{c|c|c|c}

\hline

Fitted & Experimental & Value &
$\Delta\chi^2$ \\

parameters & constraint~(\ref{17p}),\cite{xing07} &  obtained
from fit & \\

\hline

$\lambda_{\rm u}$ & $1.28^{+0.50}_{-0.43}$ &
$1.30^{+0.49}_{-0.43}$ & $0.0016$ \\

$\lambda_{\rm c}$ & $-(0.624\pm 0.083)\times
10^{3}$ & $-(0.643\pm 0.081)\times
10^{3}$ & 0.052 \\

$\lambda_{\rm t}$ & $(172.5\pm 3.0)\times
10^{3}$ & $(172.4\pm 3.0)\times
10^{3}$ & 0.0011 \\

$\lambda_{\rm d}$ & $2.91^{+1.24}_{-1.20}$ &
$2.79^{+0.39}_{-0.38}$ & $0.010$ \\

$\lambda_{\rm s}$ & $-(55^{+16}_{-15})$ &
$-(35.7\pm 4.6)$ & $1.66$ \\

$\lambda_{\rm b}$ & $(2.89\pm 0.09)\times
10^{3}$ & $(2.899\pm 0.090)\times
10^{3}$ & 0.010 \\

$|d'|$ & --- & $8.3^{+1.5}_{-1.1}$ & --- \\

\hline

Observable & Experiment~\cite{pdg06} &
Prediction~(\ref{vajnon}),(\ref{jvnon}) & $\Delta\chi^2$ \\

\hline

$|V_{\rm ud}|$ & $0.97383\pm
0.00024$ & $0.97385$ & $0.0069$ \\

$|V_{\rm us}|$ & $0.2272\pm
0.0010$ & $0.2272$ & $0.00$ \\

$|V_{\rm cd}|$ & $0.2271\pm
0.0010$ & $0.2269$ & $0.040$ \\

$|V_{\rm cs}|$ & $0.97296\pm
0.00024$ & $0.97288$ & $0.11$ \\

$J(V)$ & $(3.08\pm 0.18)\times 10^{-5}$ &
$3.08\times 10^{-5}$ & $0.00$ \\

\end{tabular}

\end{ruledtabular}

\end{table}

\begin{table}

\caption{Fritzsch- and CGS-type mass matrices in physical basis.
Experimentally observed, predicted, and $\Delta\chi^2$ values of the
derived parameters $a$, $b$, $c$, and $a'$ and $c'$, of
Eqs.~(\ref{15bnon}) and (\ref{15bpnon}) in the up and down quark
sectors, respectively. The predicted value for $b'$ of
Eqs.~(\ref{15bpnon}) is also displayed. Experimental and predicted
values were determined by using the entries of Table~\ref{table1} for
the experimental quark masses constraints and the fitted values of the
eigenvalues, respectively. \label{table5b} }

\begin{ruledtabular}

\begin{tabular}{c|c|c|c}

Derived & Experiment & Prediction & $\Delta\chi^2$ \\

parameters & (MeV) & (MeV) & \\

\hline

$a$ & $28.3^{+5.8}_{-5.1}$ & $28.9$ & 0.011\\

$b$ & $(10.36\pm 0.70)\times 10^{3}$ & $10.52\times 10^{3}$ &
0.052 \\

$c$ & $(171.9\pm 3.0)\times 10^{3}$ & $171.8\times 10^{3}$ &
0.0011 \\

$a'$ & $12.8^{+3.3}_{-3.2}$ & $10.0$ & 0.76 \\

$c'$ & $(2.838\pm 0.091)\times 10^{3}$ & $2.866\times 10^{3}$ &
0.095 \\

\hline

$b'$ & --- & $309$ & --- \\

\end{tabular}

\end{ruledtabular}

\end{table}

\begin{table}

\caption{Comparison of the $\chi^2/{\rm (dof)}$ for the various types
of mass matrices and quark bases considered using as
experimental constraints the determined values of
the quark masses at the $M_Z$ energy scale, the magnitudes of
the quark mixing matrix elements $V_{\rm ud}$, $V_{\rm us}$,
$V_{\rm cd}$, and $V_{\rm cs}$, and the Jarlskog invariant
$J(V)$. Note that the number of parameters comes from the
non-zero elements of the two mass matrices. These are 3
parameters from a diagonal mass matrix plus those in the other
non-diagonal mass matrix. There are 11 summands in the $\chi^2$
function as explained at the end of
Sec.~\ref{notation}.\label{table6} }

\begin{ruledtabular}

\begin{tabular}{l|c|c|c|c}

Table & Type of & Basis & Number of & $\chi^2/{\rm (dof)}$ \\

& mass matrix &  & parameters & \\

\hline

I, II & Fritzsch & Physical ($\phi_{A'}$ and $\phi_{B'}$
free) & 8 & $4.23/3=1.41$ \\

III, IV & & Physical ($\phi_{A'}=-\pi/2$, $\phi_{B'}=0$)
& 6 & $4.84/5=0.97$ \\

\hline

V, VI & Stech & $M_{\rm u}$ diagonal & 7 & $9.10/4=2.28$
\\

\hline

VII, VIII & CGS & $M_{\rm d}$ diagonal & 7 &
$5.92/4=1.48$  \\

IX, X & &$M_{\rm u}$ diagonal & 7 & $15.50/4=3.88$ \\

\hline

XI, XII & $M_{\rm u}$ Fritzsch-type & & &  \\
& and $M_{\rm d}$ CGS-type & Physical & 7 & $1.89/4=0.47$ \\

\end{tabular}

\end{ruledtabular}

\end{table}

\begin{table}

\caption{Comparison of the $\chi^2/{\rm (dof)}$ for the various types
of mass matrices and quark bases considered using as experimental
constraints the determined values of the quark masses at the
2~GeV energy scale ($m_{\rm u}=2.2^{+0.8}_{-0.7}\,\,{\rm MeV}$,
$m_{\rm d}=5.0\pm 2.0\,\,{\rm MeV}$, $m_{\rm s}=95\pm
25\,\,{\rm MeV}$, $m_{\rm c}=1.07\pm 0.12\,\,{\rm GeV}$, $m_{\rm
b}=5.04^{+0.16}_{-0.15}\,\,{\rm GeV}$, $m_{\rm
t}=318.9^{+13.1}_{-12.3}\,\,{\rm GeV}$)~\cite{xing07}, the
magnitudes of the quark mixing matrix elements $V_{\rm ud}$,
$V_{\rm us}$, $V_{\rm cd}$, and $V_{\rm cs}$, and the Jarlskog
invariant $J(V)$. Note that the number of parameters comes from
the non-zero elements of the two mass matrices. These are 3
parameters from a diagonal mass matrix plus those in the other
non-diagonal mass matrix. There are 11 summands in the $\chi^2$
function as explained at the end of
Sec.~\ref{notation}.\label{table6-2gev} }

\begin{ruledtabular}

\begin{tabular}{l|c|c|c|c}

Table & Type of & Basis & Number of & $\chi^2/{\rm (dof)}$ \\

& mass matrix &  & parameters & \\

\hline

I, II & Fritzsch & Physical ($\phi_{A'}$ and $\phi_{B'}$
free) & 8 & $4.80/3=1.60$ \\

III, IV & & Physical ($\phi_{A'}=-\pi/2$, $\phi_{B'}=0$)
& 6 & $5.49/5=1.10$ \\

\hline

V, VI & Stech & $M_{\rm u}$ diagonal & 7 & $11.00/4=2.75$
\\

\hline

VII, VIII & CGS & $M_{\rm d}$ diagonal & 7 &
$5.39/4=1.35$  \\

IX, X & &$M_{\rm u}$ diagonal & 7 & $17.99/4=4.50$ \\

\hline

XI, XII & $M_{\rm u}$ Fritzsch-type & & &  \\
& and $M_{\rm d}$ CGS-type & Physical & 7 & $2.47/4=0.62$ \\

\end{tabular}

\end{ruledtabular}

\end{table}


\begin{thebibliography}{}
\bibitem{footnote}
It is interesting to note that for quark
masses given in~\cite{pdg06}, namely, $m_{\rm u}(2~{\rm
GeV})=1.5\,\,{\rm to}\,\,3.0\,\,{\rm MeV}$, $m_{\rm d}(2~{\rm
GeV})=3\,\,{\rm to}\,\,7\,\,{\rm MeV}$, $m_{\rm s}(2~{\rm
GeV})=95\pm 25\,\,{\rm MeV}$, $m_{\rm c}(m_{\rm c})=1.25\pm
0.09\,\,{\rm GeV}$, $m_{\rm b}(m_{\rm b})=4.20\pm 0.07\,\,{\rm
GeV}$, and $m_{\rm t}=174.2\pm 3.3\,\,{\rm GeV}$, the
phenomenological conclusion of our analysis is the
same~\cite{version1}. This would suggest that the role of the
small masses (since they evolve slowly) is more important for
our analysis.
\end{thebibliography}
\end{document}